\documentclass[showpacs,preprintnumbers,amsmath,amssymb,double]{revtex4}
\usepackage{graphicx}
\usepackage{dcolumn}
\usepackage{bm}

\begin{document}
\title{Influence of chemical and magnetic interface properties of Co-Fe-B / MgO / Co-Fe-B tunnel junctions on the annealing temperature dependence of the magnetoresistance}
\author{J. Schmalhorst}
\email{jschmalh@physik.uni-bielefeld.de}
\author{A. Thomas}
\author{G. Reiss}
\affiliation{Thin Films and Nano Structures, Department of Physics, Bielefeld University, 33501 Bielefeld, Germany}
\author{X. Kou}
\affiliation{Lanzhou University, 222 South Tianshui Road, Lanzhou 7300000, China}
\author{E. Arenholz}
\affiliation{Lawrence Berkeley National Laboratory, Berkeley, CA 94720, USA}
\date{\today}
\begin{abstract}
The knowledge of chemical and magnetic conditions at the Co$_{40}$Fe$_{40}$B$_{20}$ / MgO interface is important to interpret the strong annealing temperature dependence of tunnel magnetoresistance of Co-Fe-B / MgO / Co-Fe-B magnetic tunnel junctions,  which increases with annealing temperature from 20\% after annealing at 200$^{\circ}$C up to a maximum value of 112\% after annealing at  350$^{\circ}$C. While the well defined nearest neighbor ordering indicating crystallinity of the MgO  barrier does not change by the annealing,  a small amount of interfacial Fe-O at the lower Co-Fe-B / MgO interface is found in the as grown samples, which is completely reduced after annealing at 275$^{\circ}$C. This is accompanied by a simultaneous increase of the Fe  magnetic moment and the tunnel magnetoresistance. However, the TMR of the MgO based junctions increases further for higher annealing temperature which can not be caused by Fe-O reduction. The occurrence of an x-ray absorption near-edge structure above the Fe and Co L-edges after annealing at 350$^{\circ}$C indicates the recrystallization of the Co-Fe-B electrode. This is prerequisite for coherent tunneling and has been suggested to be responsible for the further increase of the TMR above 275$^{\circ}$C. Simultaneously, the B concentration in the Co-Fe-B decreases with increasing annealing temperature, at least some of the B diffuses towards or into the MgO barrier and forms a B$_2$O$_3$ oxide. 
\end{abstract}
\maketitle

Magnetic tunnel junctions (MTJs) consisting of ferromagnetic electrodes separated by a thin insulating tunnel barrier are basic elements for spintronic devices such as programmable logic devices\cite{rei05b}, magnetic sensors and nonvolatile memories\cite{tre01}. MTJs with MgO barrier have attracted increasing attention due to the theoretically predicted and experimentally verified huge tunneling magnetoresistance (TMR) effect\cite{GRIV23,GRIV24}. By using initially amorphous Co-Fe-B electrodes a room temperature TMR of more than 350\% has been achieved\cite{ohn05}. Coherent spin-polarized tunneling has been suggested to be responsible for the high TMR effect\cite{GRIV24,SY99,SY100}.

In this work we utilized x-ray absorption spectroscopy (XAS) and x-ray magnetic circular dichroism (XMCD) to investigate the chemical and magnetic properties at the lower Co$_{40}$Fe$_{40}$B$_{20}$ / MgO interface of our Co-Fe-B / Mg interlayer / MgO / Co-Fe-B MTJs. 
The MTJs were prepared in a magnetron sputtering system with a base pressure of $1\times 10^{-7}$mbar. The layer stack consists of Ta$^{5nm}$ / Cu$^{30nm}$ / Ta$^{5nm}$ / Cu$^{5nm}$ / Mn$_{83}$Ir$_{17}^{12nm}$ / Co$_{40}$Fe$_{40}$B$_{20}^{4nm}$ / Mg$^{0.5\mbox{ or }0.75nm}$ / MgO$^{1.5nm}$ / Co$_{40}$Fe$_{40}$B$_{20}^{6nm}$ / Ta$^{5nm}$ / Cu$^{40nm}$ on thermally oxidized (50 nm) silicon (100) wafers. The MgO was RF-sputtered from a stoichiometric MgO target. The MTJs were vacuum annealed for 60 minutes at temperatures ranging from 200$^{\circ}$C to 450$^{\circ}$C. A magnetic field of 1 kOe was applied during annealing to achieve the exchange bias of the Mn-Ir / Co-Fe-B double layer. Then, the MTJs were covered by 30nm Au and patterned by optical lithography and Ar$^+$-ion beam etching. The junction area ranges from $2\times 2\mu \mbox{m}^2$ to $300\times 300\mu \mbox{m}^2$. The transport properties of the MTJ were measured by using conventional 2-probe method with a constant dc bias voltage. 
For probing the chemical and magnetic properties at the lower Co-Fe-B / Mg-O interface, we grew additional stacks similar to the full MTJ stacks described above, but stopped the deposition after forming the MgO barrier. The stacks with Mg interlayer thickness of 0.5nm (0.75nm) were annealed at 275$^{\circ}$C and 350$^{\circ}$C (325$^{\circ}$C) under ultra-high vacuum condition (with applied magnetic field of 1 kOe). XAS and  XMCD were performed at beamline 4.0.2 of the Advanced Light Source, Berkeley, USA. The $L$-edges of Co and Fe and the B $K$-edge were studied. Surface-sensitive total electron yield spectra (TEY)\cite{kao94}  were recorded, the x-rays angle of incidence was 30$^o$ to the sample surface. XMCD spectra were obtained by applying a magnetic field (max. $\pm$ 0.55T) along the x-ray beam direction using elliptically polarized radiation with a  polarization of 90\%. XAS intensity and XMCD effect are defined as $(I^+ + I^-)/2$ and ($I^+-I^-$), respectively, where $I^+$ ($I^-$) denote spectra for parallel (antiparallel) orientation of photon spin and magnetic field. 

\begin{figure}
\begin{center}
\includegraphics[height=6cm]{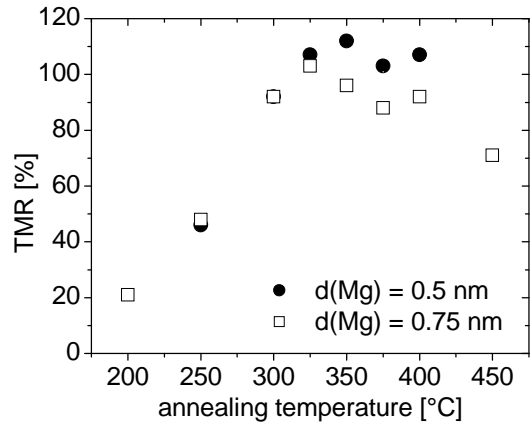}
\caption{Annealing temperature dependence of the TMR for two Mg interlayer thicknesses $d(Mg)$ measured at room temperature with 10mV bias voltage}
\label{fig:1}
\end{center}
\end{figure}
As shown in Fig. \ref{fig:1} the TMR increases strongly with annealing temperature for both Mg interlayer thickness. Starting at around 20\% after annealing at 200$^{\circ}$C it reaches a maximum value of 112\% after annealing at 350$^{\circ}$C, while the TMR amplitude above 325$^{\circ}$C is generally slightly larger for an Mg interlayer thickness of 0.5nm than for 0.75nm. 

\begin{figure}
\begin{center}
\includegraphics[height=6cm]{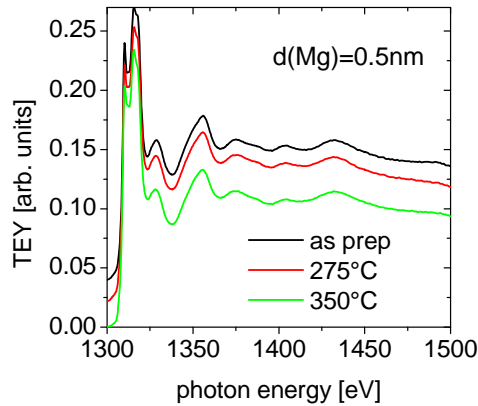}
\caption{XAS spectra at the Mg $K$-edge for an Mg interlayer thickness of 0.5nm and different annealing temperatures. Similar results were found for $d(Mg)= 0.75$nm.}
\label{fig:2}
\end{center}
\end{figure}
\begin{figure}
\begin{center}
\includegraphics[height=8cm]{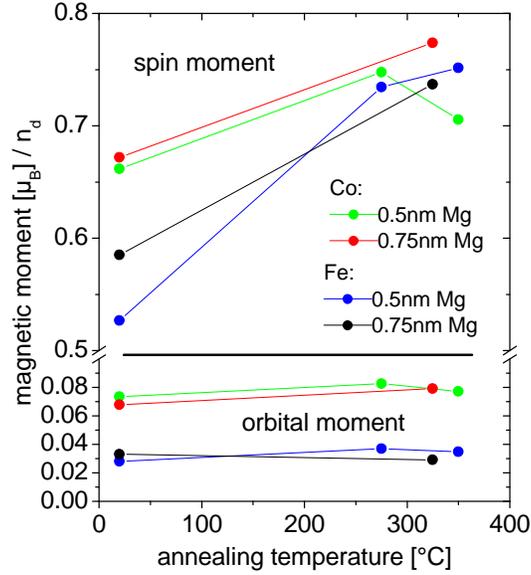}
\caption{Spin and orbital magnetic moments per number of 3d-holes, $n_d$, of Fe and Co at the lower Co-Fe-B / Mg-O interface 
for an Mg interlayer thickness of 0.5nm and 0.75nm as a function of the annealing temperature. The magnetic moments were extracted 
from the  XMCD and XAS data by means of the XMCD sum-rules\cite{set95}.}
\label{fig:4}
\end{center}
\end{figure}

The x-ray absorption near-edge structure (XANES) at the Mg K-edge (Fig. \ref{fig:2}) indicates a well defined nearest neighbor ordering typical for a crystalline MgO phase\cite{mur95} already in the as prepared state. The spectra and, therefore, the local order of the MgO remains unchanged after annealing. This indicates that the metallic Mg interlayer was nearly completely oxidized during the RF-sputter deposition of the MgO layer.  This fits well to the fact that a small amount of Fe-O was found below the barrier in the as prepared samples as discussed below. 
\begin{figure}
\begin{center}
\includegraphics[height=12cm]{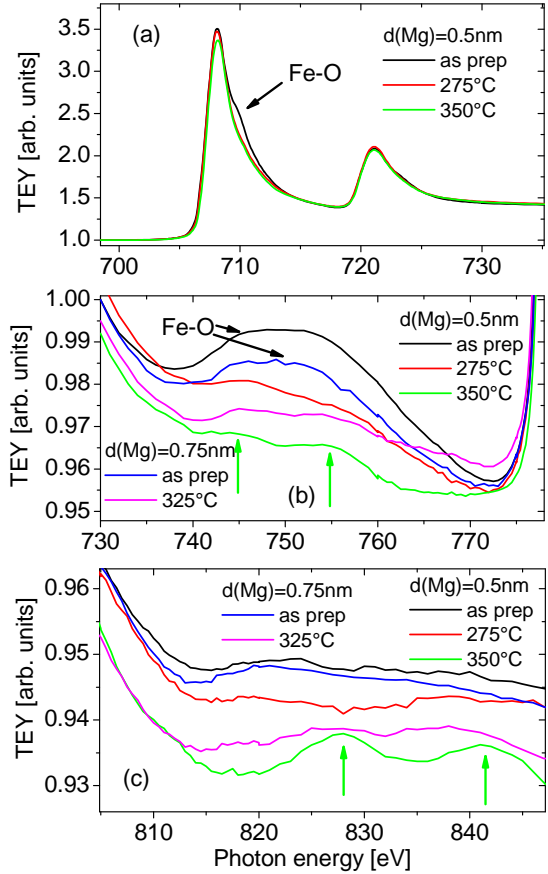}
\caption{(a) XAS spectra at the Fe $L$-edge for an Mg interlayer thickness of 0.5nm and different annealing temperatures. The shoulder 
above the $L_3$-resonance indicating Fe-O was also observed for the as prepared sample with $d(Mg)=0.75$nm.
 XANES above the (b) Fe and the (c) Co $L$-edge for Mg interlayer thickness of 0.5nm and 0.75nm and different annealing temperatures.}
\label{fig:3}
\end{center}
\end{figure}
The Fe-O is identified by a typical shoulder just above the $L_3$-resonance\cite{whi01,rei04c} (see Fig. \ref{fig:3}a) and also results in a very broad typical XANES structure around a photon energy of 750eV (see Fig. \ref{fig:3}b). As one would expect, both Fe-O features are larger for the as prepared samples with only $d(Mg)=0.5$nm Mg interlayer thickness.
After annealing at 275$^{\circ}$C the Fe-O shoulder vanished (Fig. \ref{fig:3}a) completely indicating a reduction of the oxide to metallic Fe. This behavior is similar to what we found for Al-O barrier based MTJs with Co-Fe electrode\cite{rei04c} and accompanied by a considerable increase of the TMR amplitude and of the interfacial Fe magnetic moment (Fig. \ref{fig:4}).
The XAS spectra at the Co $L_{2,3}$-resonances (not shown here) do not hint to any Co-O formation at the interface during the deposition process. However, in contrast to the Al-O based junctions\cite{rei04c} the TMR of our MgO based MTJs increases further above 275$^{\circ}$C, which can not result from further Fe-O reduction. A new spectral feature, however, occurs at higher annealing temperature: for Fe (Fig. \ref{fig:3}b) as well as for Co (Fig. \ref{fig:3}c) clear XANES oscillations above the Fe and Co L-edges are observed indicating the recrystallisation of the Co-Fe-B electrode. The oscillations are strongest for the highest annealing temperature of 350$^{\circ}$C (see green arrows in Fig. \ref{fig:3}b and c). A recrystallization of the Co-Fe-B electrodes being amorphous in the as prepared state is essential for the interpretation of the increasing TMR based on coherent tunneling, because a coherent or "quasi-epitaxial" electrode / barrier interface is required\cite{GRIV24,SY99,SY100}.
Therefore, our results confirm that the further increase of the TMR above 275$^{\circ}$C results from an enhanced coherent tunneling. 

\begin{figure}
\begin{center}
\includegraphics[height=7cm]{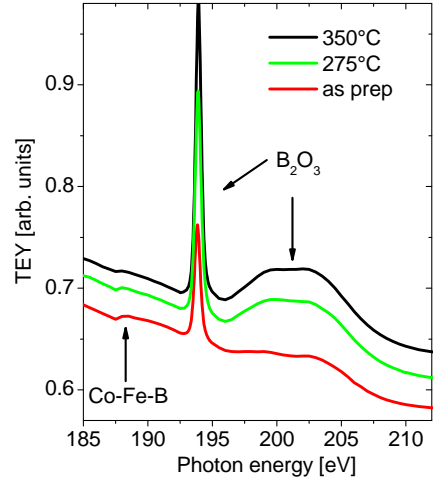}
\caption{XAS spectra at the B $K$-edge for an Mg interlayer thickness of 0.5nm and different annealing temperatures.}
\label{fig:5}
\end{center}
\end{figure}
Finally, we like to note, that the boron concentration in the Co-Fe-B electrode decreases with increasing annealing temperature. 
As can be seen from Fig. \ref{fig:5}, the edge jump corresponding to (metallic) boron in the Co-Fe-B (marked by an arrow) is reduced with increasing annealing temperature. On the other hand, the narrow peak at 194eV and the very broad peak around 201eV, both are fingerprints of B$_2$O$_3$\cite{rol91},  become much stronger with higher annealing temperature. This indicates, that boron-oxide is accumulated in the MgO barrier or close to the electrode / barrier interface. Although the influence of the boron or its oxides on the coherent tunneling process and, therefore, on the TMR amplitude is not clear  so far, one might speculate, that this interfacial imperfection should reduce rather than increase the TMR in our junctions. Recently, Burton {\it et al.}\cite{CoFeBtheo} found a detrimental influence of B at the interface to the MgO barrier by using {\it First principle calculations} and Santos and Moodera\cite{moo07} found a vanishing spinpolarization in Co / B$_2$O$_3$ / SC-Al and Fe / B$_2$O$_3$ / SC-Al junctions. 

In summary, we reported on chemical and magnetic properties at the lower electrode / barrier interface in Co-Fe-B / MgO / Co-Fe-B MTJs. The increase of the TMR for annealing temperatures up to about 275$^{\circ}$ is correlated to a thermally induced reduction of interfacial Fe-O accompanied by a strong increase of the Fe magnetic spin moment. Similar results have been found for Al-O based MTJs previously. The further TMR increase up to a maximum value of 112\% after annealing at  350$^{\circ}$C is suggested to result from a completely different mechanism, namely from an enhanced contribution of coherent 
tunneling. Finally, an accumulation of B-oxide in or just below the MgO barrier was found, which occurs as important reason for the current limitation of the TMR amplitude in our Co-Fe-B / MgO / Co-Fe-B MTJs. 

Financial support by the Deutsche Forschungsgemeinschaft (DFG) is  gratefully acknowledged. The Advanced Light Source is supported by the U.S. Department of Energy under Contract No. DE-AC03-76SF00098.

%\bibliography{BibJS}

\end{document}